\documentclass[sigconf]{acmart} %
\AtBeginDocument{%
  }

\setcopyright{acmlicensed} %
\copyrightyear{2026}
\acmYear{2026}
\setcopyright{cc}
\setcctype{by}
\acmConference[CCS '26]{Proceedings of the 2026 ACM SIGSAC Conference on Computer and Communications Security}{November 15--19, 2026}{The Hague, Netherlands}
\acmBooktitle{Proceedings of the 2026 ACM SIGSAC Conference on Computer and Communications Security (CCS '26), November 15--19, 2026, The Hague, Netherlands}
\acmDOI{10.1145/3830454.3846563}
\acmISBN{979-8-4007-2871-6/2026/11}

\usepackage{tabularx}
\usepackage{svg}
\usepackage{booktabs}
\usepackage{color}
\usepackage[dvipsnames]{xcolor} %
\usepackage{totpages} %
\usepackage{graphicx}
\usepackage[labelformat=simple]{subcaption}
\usepackage{xspace}
\usepackage{multirow}
\usepackage{makecell} %
\usepackage{pifont} %
\usepackage{array}
\usepackage{longtable}
\usepackage[nolist,nohyperlinks]{acronym}
\usepackage{enumitem} %
\usepackage[ruled,vlined]{algorithm2e}
\usepackage{float} %
\usepackage{placeins} %
\usepackage{cuted} %
\usepackage{pgfplots}
\usepackage{pgfplotstable}
\usepackage{tikzviolinplots}
\definecolor{tgaShadeA}{RGB}{23,32,66}
\definecolor{tgaShadeB}{RGB}{47,94,141}
\definecolor{tgaShadeC}{RGB}{69,151,163}
\definecolor{tgaShadeD}{RGB}{118,189,175}
\definecolor{tgaShadeE}{RGB}{203,229,197}
\definecolor{tgaShadeF}{RGB}{250,250,210}
\pgfplotscreateplotcyclelist{tgaCycle}{
    {draw=tgaShadeA, fill=tgaShadeA!55!white},
    {draw=tgaShadeB, fill=tgaShadeB!55!white},
    {draw=tgaShadeC, fill=tgaShadeC!60!white},
    {draw=tgaShadeD, fill=tgaShadeD!70!white},
    {draw=tgaShadeE, fill=tgaShadeE!80!white},
    {draw=tgaShadeF, fill=tgaShadeF!85!white}
}
\pgfplotscreateplotcyclelist{tgaLineCycle}{
    {draw=tgaShadeA},
    {draw=tgaShadeB},
    {draw=tgaShadeC},
    {draw=tgaShadeD},
    {draw=tgaShadeE},
    {draw=tgaShadeF}
}
\pgfplotsset{
    compat=1.18,
    filter discard warning=false,
    cycle list name=tgaCycle,
    colormap={tgaHeat}{
        rgb255(0cm)=(23,32,66)
        rgb255(1cm)=(47,94,141)
        rgb255(2cm)=(69,151,163)
        rgb255(3cm)=(146,207,184)
        rgb255(4cm)=(250,250,210)
    },
    colormap={tgaHeatRev}{
        rgb255(0cm)=(23,32,66)
        rgb255(1cm)=(47,94,141)
        rgb255(2cm)=(69,151,163)
        rgb255(3cm)=(146,207,184)
        rgb255(4cm)=(250,250,210)
    },
}
\pgfplotsset{
    tgaPlotBase/.style={
        axis line style={semithick, color=black!70},
        tick style={color=black!70},
        label style={font=\footnotesize},
        tick label style={font=\scriptsize},
        title style={font=\small},
        legend style={font=\scriptsize}
    },
    tgaBarAxis/.style={
        tgaPlotBase,
        scale only axis,
        ymajorgrids=true,
        grid style={dotted, color=black!25},
        enlarge x limits=0.02,
        bar width=6pt,
        cycle list name=tgaCycle
    },
    tgaDistributionAxis/.style={
        tgaPlotBase,
        scale only axis,
        xbar stacked,
        xmin=0,
        xmax=1,
        y dir=reverse,
        xmajorgrids=true,
        grid style={dotted, color=black!25},
        cycle list name=tgaCycle
    },
    tgaHeatmapAxis/.style={
        tgaPlotBase,
        axis on top,
        scale only axis,
        enlargelimits=false,
        colormap name=tgaHeat,
        colorbar style={tick label style={font=\scriptsize}}
    }
}
\usepgfplotslibrary{groupplots}
\usepgfplotslibrary{colormaps}
\usepgfplotslibrary{fillbetween}
\usetikzlibrary{arrows.meta,positioning,shapes.multipart,shapes.geometric,fit,calc,matrix,fillbetween}

\usepackage{ulem}
\newcommand{\etal}{et~al.\xspace}
\newcommand{\eg}{e.g.,\xspace}
\newcommand{\ie}{i.e.,\xspace}

\newcommand{\vsix}{IPv6\xspace}
\newcommand{\icmpvsix}{ICMPv6\xspace}
\newcommand{\vfour}{IPv4\xspace}

\newcommand{\zmap}{ZMap\xspace}

\newcommand{\yarrpsix}{Yarrp6\xspace}

\ifdefined\isFinalized
\newcommand{\NewCommentType}[3]{}
\else
\newcommand{\NewCommentType}[3]{\expandafter\newcommand\csname #1\endcsname[1]{{\color{#2}{#3: ##1}} }}
\fi

\usepackage[override]{cmtt} %

\graphicspath{{figures/}}
\newcommand{\pool}{NTP Pool\xspace}
\newcommand{\punkt}[1]{\item\textbf{#1}:}
\providecommand{\e}[1]{\ensuremath{\times 10^{#1}}}

\begin{document}

\title[Cleaning the NTP Pool: Detecting and Mitigating NTP-Sourced IPv6 Scanning]{Cleaning the NTP Pool:\\Detecting and Mitigating NTP-Sourced IPv6 Scanning}

\author{Erik Rye}
\affiliation{%
  \institution{Johns Hopkins University}
  \city{}
  \state{}
  \country{}
}
\email{rye@jhu.edu}

\author{Robert Beverly}
\affiliation{%
  \institution{San Diego State University}
  \city{}
  \state{}
  \country{}
}
\email{rbeverly@sdsu.edu}

\begin{abstract} 
The ephemeral and random nature of IPv6 client addresses presents a
practical challenge to attacks that depend on Internet-wide
scanning or reconnaissance -- the adversary must first \emph{find} the
client's IPv6 address.  While a well-positioned passive adversary can
potentially harvest some active IPv6 client addresses, such power is
typically reserved for \eg large CDNs or Internet exchange points.  In
contrast, prior work has shown the ease with which a low-power
entity can join the volunteer-based \pool and harvest
large quantities of active IPv6 client addresses.

In this work, we develop a methodology to not only rigorously identify such 
IPv6 address harvesting and the entities gathering addresses, 
but also characterize \emph{what} these entities subsequently do
with the addresses.  Specifically, we query all \pool servers 
across the global Internet over
the course of one year using unique IPv6 client addresses, and 
monitor and correlate any later activity targeting these addresses.
In sum, we identify 22 \pool servers, within 4 primary clusters, 
that are part of larger monitoring infrastructures that utilize
the gathered addresses for reconnaissance, port scanning, 
and service and vulnerability enumeration.  To better 
understand the legal and ethical gray area of such behavior, we
engage with both the \pool operators and a cybersecurity insurance firm running
one of the harvesting and scanning clusters.  We are in discussions with the
\pool to integrate our system into their monitoring
infrastructure to remove such NTP servers, and the cybersecurity insurance firm
    changed its operational policy to be more transparent and
provide clear opt-out mechanisms.  

\end{abstract}



\maketitle

\section{Introduction}
\label{sec:intro}

Core Internet infrastructure has a long history of employing
distributed control with implicit trust; notable examples include
BGP and DNS.  Indeed, core infrastructure owners and providers are
often well-positioned to mount both active and passive attacks against
users, applications, and traffic that depend on that infrastructure.
In this work, we examine a way in which the deployment of one such
core Internet infrastructure component -- the Network Time Protocol
(NTP) \cite{rfc5905} -- is being exploited in an unintended manner, 
specifically to harvest IPv6 client addresses that would otherwise
be unguessable, and subsequently port scan them.

Continually, and transparently to users, Internet devices use
NTP to provide accurate time.  While a number of operating system and
mobile phone vendors operate their own NTP infrastructure, the \pool
is widely used by embedded Linux and IoT devices.  For instance,
Beverly and Rye estimate that the \pool consists of over 15k servers
and its infrastructure receives over 100k DNS queries per
second~\cite{beverly2026ntpfool}, while Moura~\etal found
that the \pool was the most popular NTP deployment as measured at
a DNS root (accounting for 71\% of all queries)~\cite{moura2024deep}.  Unique among most Internet
infrastructure, the \pool is volunteer based and operationally
distributed -- thus \emph{anyone} can operate an NTP server and easily join
the \pool.  To receive NTP queries from real IPv6 clients, the barrier to entry 
is simply installing and running an NTP server, \eg on a VPS, and 
registering its IP address with the \pool.  

While significant prior work has examined potential time attacks that
leverage this volunteer-based model~\cite{malhotra2015attacking,
annessi2017s, rytilahti2018masters, czyz2014taming,
deutsch2018preventing, perry2021devil,beverly2026ntpfool}, we identify
a new form of passive attack: NTP servers within the \pool that gather
client \vsix addresses to enable subsequent service and port scanning
-- a phenomenon we term ``back-scanning.''  As compared to \vfour, a
fundamental difficulty in scanning and enumerating hosts and services
within \vsix is the massive address space.  \vsix clients typically
use Privacy Extensions (PE), where the lower 64 bits of the address,
the Interface Identifier (IID), are both random and ephemeral.  Thus,
passively gathering NTP queries from \vsix clients provides a ready
source of valid and operational \vsix addresses that would otherwise
be impossible to find.  

Toward understanding, characterizing, and mitigating \vsix back-scanning
enabled by the \pool, this work makes the following contributions:

\begin{enumerate}
  \item Development of a rigorous methodology to 
  detect \pool back-scanning, and global deployment
  to induce and characterize back-scanning, \eg 
  sampling rates, scan delays, etc.
  \item Identification of 4 clusters of back-scanners
  and a deeper understanding of their intent.
  \item Coordination with the \pool operators to 
  integrate our system into their infrastructure to 
  remove \vsix-harvesting NTP servers from the \pool.
\end{enumerate}

The remainder of this paper is organized as follows.  We review
prior work on IPv6 address discovery, scanning, and the \pool
in~\S\ref{sec:background}.  \S\ref{sec:methodology} details our
methodology for identifying members of the \pool harvesting addresses
and performing back-scanning.  We cluster and characterize the 
\pool-initiated scans in \S\ref{sec:results}, and delve into 
the application-layer intent of the scans in \S\ref{sec:casestudy}.
Toward impact, in \S\ref{sec:validation} we describe our engagement with the \pool 
operators to make the system more robust, and detail interactions with
a major scan operator that resulted in better operational
transparency.  Finally, we provide long-term recommendations
in~\S\ref{sec:recommendations} and conclude in~\S\ref{sec:concl}.

\section{Background and Related Work}
\label{sec:background}

There is a rich history of Internet reconnaissance, including 
vertical (TCP or UDP ports) and horizontal (address) scanning, as
well as subsequent service discovery and characterization or 
exploitation.  First popularized in tools such as nmap~\cite{lyon2009nmap}, 
high-speed Internet-wide scanning was pioneered by Durumeric \etal
in \zmap~\cite{durumeric2013zmap}.  While \zmap enables exhaustive
IPv4 scanning, the massive and sparsely populated IPv6 address space
renders exhaustive enumeration techniques infeasible.

\subsection{IPv6 Address Discovery}

\begin{itemize}[wide]
\punkt{Active \vsix Scanning} Given the challenges inherent in remote IPv6 host discovery, a large
body of work emerged to identify active IPv6 targets.
Gasser combined multiple sources of data, including rDNS,
traceroutes, and passive flows, to create the first IPv6
hitlist~\cite{gasserscanning}.  Subsequent research from Borgolte
and Fiebig \etal utilized DNS to further enumerate IPv6
addresses~\cite{borgolte2018enumerating,fiebig2017something}, while
Beverly \etal synthesized new hitlists for the specific purpose of
better IPv6 topology discovery~\cite{beverly2018ip}.

Extending the idea of hitlists, Murdock \etal performed some of the
first work to generate candidate IPv6 addresses based on input
training (or ``seed'') data~\cite{murdock2017target}; the intuition
being that inherent addressing and allocation patterns are not random
and can be learned to find candidates that are more likely to be
responsive than purely random addresses.  A large body of 
follow-on work on Target Generation Algorithms (TGAs) has subsequently 
refined this basic idea with different AI/ML algorithms.  Steger \etal
provide a survey and evaluation of the major IPv6
TGAs~\cite{steger2023target}.

\punkt{Passive \vsix Discovery} An alternative approach for finding active IPv6 addresses is to
collect them \emph{passively}, \eg by monitoring a network link or
traffic arriving at some popular Internet infrastructure.  For
example, Plonka and Berger utilize traffic observed at a large
\ac{CDN} to characterize the temporal and spatial
behavior of IPv6 addresses~\cite{plonka2015temporal}. Most
closely related to the present work, Rye and Levin utilize the
\pool to collect active client IPv6 addresses~\cite{rye2023ipv6}
and maintain a regularly updated set of curated active /64s observed
through their IPv6 observatory~\cite{ipv6observ}.
\end{itemize}

\subsection{Understanding \vsix Scanning and Scanners}

In 2018, Fukuda and Heidemann used DNS backscatter from the B-root DNS server to
identify an average of 16 \vsix scanners per week~\cite{fukuda2018knocks}. As
scanner \vsix addresses will likely stimulate rDNS queries from \eg firewalls,
the presence of many rDNS queries for a specific address may indicate scanning
activity. They confirm their scanner identification through contemporaneous
backbone traffic and a darknet. In contrast with this passive scanner
identification approach, in this work, we actively query NTP servers
participating in the NTP Pool to elicit \vsix scanning.

Richter~\etal examine \vsix scanning behavior through the lens of a major
\ac{CDN}~\cite{richter2022illuminating}. From 15 months of firewall logs for
unsolicited \vsix TCP traffic for ports 80 and 443, they find that \vsix
scanning behavior varies significantly from \vfour scanning. For instance,
scanners may use an \vsix source address to probe only a single target before
choosing a new source address to obscure the scanner's identity. 

Hiesgen~\etal developed Spoki, a lightweight system to capture additional
information from Internet-wide scanners beyond the initial TCP SYN packet
captured by most network telescopes~\cite{hiesgen2022spoki}. Spoki accomplishes
this by responding to stateless scanners in order to elicit future, stateful
scanning and capture the initial data sent in a TCP connection. Unlike our
work, Spoki focuses exclusively on \vfour.

\subsection{The NTP Pool}
\label{sec:ntppool}

While often invisible and in the background for the majority of users,
the Network Time Protocol (NTP) is a critical Internet service that
provides synchronized accurate time over the Internet.  While many
major providers, \eg Google, Apple, and Microsoft, operate their own
NTP infrastructure, the large deployed base of myriad IoT and embedded
systems running Linux primarily uses the
\pool~\cite{moura2024deep,beverly2026ntpfool}.  
Servers within the \pool belong to one or more ``zones'' which 
correspond to countries and geographic regions.  The \pool operates 
a DNS infrastructure that performs coarse-grained geolocation and 
maps an incoming DNS query to a set of up to four NTP servers within 
the pool.  The \pool mapping and query assignment algorithm is
detailed in~\cite{moura2024deep}.  

Among such critical
infrastructures, the \pool is unique in that it is \emph{volunteer}
based: anyone can configure an operational NTP server, register with
the \pool, and add their server to the set of available servers
within the \pool.  Registration is trivial and only requires an email
address -- thus, entities can effectively register and join
anonymously.  As such, because of the large number of devices using the \pool, 
a volunteer NTP server in the \pool will receive a significant number
of NTP queries and afford the server operator a window into currently 
active IPv6 addresses within their zone.

\subsection{Eliciting IPv6 Scans}

Tanveer~\etal employed different indicators of network liveness -- web
crawls, open DNS resolver queries, domain registration, NTP server operation,
and Tor relay operation -- to gauge their effectiveness in generating \vsix scan
traffic and to study the behavior of \vsix scanners~\cite{tanveer2023glowing}.
Similarly, Zhao~\etal add five NTP servers to the NTP Pool to analyze the time
it takes for their servers to be scanned, as well as the type and quantity of
\vsix scan traffic that their servers attract~\cite{zhao2025exposed}.  In contrast, we investigate
whether servers in the NTP Pool are being used to collect \vsix addresses
for scanning, rather than whether operating an NTP server causes a network to be
scanned. 

Later, in 2025, Tanveer~\etal developed additional techniques to enhance
researchers' ability to attract unsolicited \vsix scan traffic to an \vsix
darknet~\cite{tanveer2025unveiling}.  These methods include issuing TLS
certificates, making BGP announcements, domain registration, and inclusion of
addresses within their network in well-known \vsix
hitlists~\cite{expanse,gasserscanning,hitlistwebsite}.  The authors partnered
with a \ac{CDN} to longitudinally analyze \vsix scanning activity over a period
of two years, which confirmed their conclusion that \vsix scanning is on the
rise.  Egloff~\etal \cite{egloff2025detailed} similarly focus on the
reactivity
of IPv6 scanning algorithms and demonstrate that scanners react
quickly to BGP prefix announcements.

Closely related to our methodology, Roberts and Plonka embed
\emph{nonces}
(a number used only once) within the lower 64 bits of IPv6 source
addresses of probes sent throughout the Internet~\cite{roberts2020watching}.  
By doing so, they can detect subsequent traffic to the nonce address
that is indicative of an entity monitoring traffic and gathering
IPv6 addresses for the purpose of scanning.  Further, because each 
probe has a unique nonce, they can identify where in the network
the monitoring is occurring.  
The authors found
that a number of entities were monitoring their traffic by observing rDNS
queries for, and port and protocol scanning directed at, their nonce addresses.
Similar to this work, we employ parts of \vsix addresses as nonces in
order to 
detect network monitoring. In contrast, we focus on understanding
whether network services, such as NTP, are used to gather IPv6
addresses and seed active scanning.

\section{Methodology}
\label{sec:methodology}

\begin{figure}[t!]
        \centering
        \resizebox{0.9\columnwidth}{!}{\includegraphics{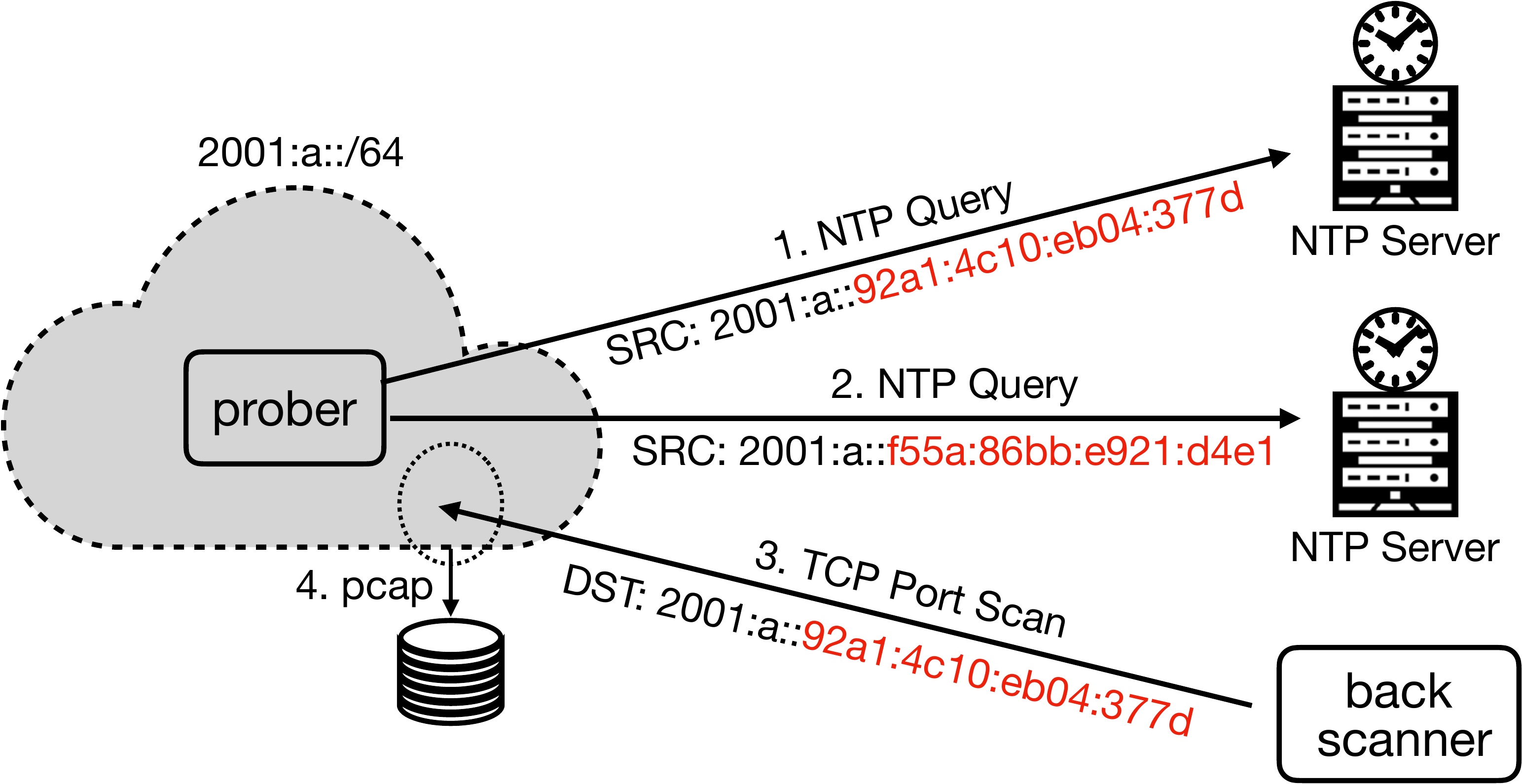}}
        \caption{Methodology: we send
         standard and compliant NTP queries to all IPv6 \pool servers
         (1).
         Each query encodes a unique nonce within the lower-64 bits
         of the source IPv6 address (2).  If an NTP server is 
         using incoming queries to gather IPv6 client addresses, 
         \eg for back-scanning (3), we detect this by capturing 
         and correlating all
         packets arriving at our vantage (4).}
        \label{fig:method}
\end{figure}

In this section, we describe our system for detecting \vsix monitoring by
NTP Pool server operators. Additionally, we describe steps taken to reduce the
likelihood of false positives, \ie incorrectly ascribing responsibility for
\vsix scanning to NTP servers, \eg due to on-path middleboxes or
network taps.  Figure~\ref{fig:method} provides a visual overview of
our methodology.

\subsection{Definitions}

We first define the following terms for clarity and consistency throughout the
remainder of this work:

\textbf{Prober:} Our host issuing NTP client request queries that
embed nonces in the source IPv6 address. In our experiments,
this is an Ubuntu 24.04 VPS located in the US;
we draw nonce addresses from the /64 network
assigned by our VPS provider. However, a larger or smaller subnet might also be
used, so long as the nonce addresses will never be used for other traffic.

\textbf{Nonce address:} A random IPv6 source address drawn from the prober's
network, used for exactly one (target, \vsix Hop Limit) pair. For instance,
\texttt{2001:a::92a1:4c10:eb04:377d} is drawn randomly from
\texttt{2001:a::/64}. We log the NTP Pool server destination,
initial Hop Limit used, and time that each nonce address probe is transmitted
for later correlation.

\textbf{Target NTP server:} An NTP server to which the prober sends NTP
requests. In this work, we collected 2,335 \vsix NTP servers operating in the
\pool via repeated DNS queries to the \pool. Prior work employed this
technique to retrieve an approximation of the total number of servers in the NTP
Pool~\cite{moura2024deep}; however, servers that opt to receive little or no NTP
traffic from the NTP Pool are statistically less likely to be discovered via
repeated DNS queries.

\textbf{Back-scan:} An unsolicited inbound packet whose destination address is a
nonce address previously used by the prober as the source of an NTP
request. We consider any traffic to 
the nonce addresses to be a back-scan, regardless of transport-layer protocol.
We log back-scans by running \texttt{tcpdump} on our prober. We ensure that we
receive traffic addressed to nonce addresses by responding to Neighbor Discovery
Protocol (NDP) Neighbor Solicitation messages for any address within our
assigned /64. We use \texttt{ndppd}~\cite{ndppd} for this purpose.

\textbf{Back-scanner:} A host that sends back-scan traffic. We make the
distinction between two different types of back-scanners: i) \emph{on-path}
back-scanners, which scan nonce addresses that did not reach the NTP server
target, and ii) \emph{endpoint} back-scanners, which scan only addresses that
reached the NTP server destination. The point of this differentiation is to
distinguish between monitoring NTP servers and monitors between the prober and
a target NTP server.

\subsection{Stimulating NTP Monitoring}

From February 2025 to April 2026 we conducted experiments to detect NTP
Pool-based \vsix monitoring from a US vantage point. Our vantage point resided
in a /64 \vsix network from which it could draw arbitrary \vsix addresses, as
only the vantage point and its gateway router were assigned \vsix addresses from
this network. 

\subsubsection{Active NTP Pool Server Collection}

We first gathered active \vsix NTP Pool servers by repeatedly querying the
NTP Pool's DNS servers. The NTP Pool assigns each NTP server to a country-level
``zone'', and responds with NTP servers for DNS queries for each country zone
using the format \texttt{<1-4>.<CC>.pool.ntp.org}, where \texttt{CC} is an ISO
3166-1 alpha-2 country code. \vsix AAAA records are returned only for the
\texttt{2.<CC>.pool.ntp.org} domain name.

The NTP Pool returns volunteer-run NTP servers from each zone with a probability
correlated with an operator-configured
``netspeed''~\cite{moura2024deep,beverly2026ntpfool}; in general, operators that
choose a higher netspeed value will have their server returned more frequently
than those with a lower netspeed from within the same country zone. Therefore, we queried each \vsix NTP Pool country
zone repeatedly over the course of a day until we had either exhausted the
number of \vsix servers advertised for that country by the NTP Pool, or until
the day had elapsed. This produced a total of 2,335 NTP servers, which we
verified responded to NTP requests. 

\subsubsection{Longitudinal NTP Queries}
\label{sec:ntpqueries}

For each of the 2,335 active \vsix NTP servers we learned from the NTP Pool DNS,
we began sending iterative NTP requests designed to both detect endpoint NTP
monitoring as well as detect and exclude on-path monitoring from our analyses. 

To do this, we progressively increased the \vsix Hop Limit for each NTP server
destination from a starting value of 1 up to 32, terminating early when we
received an NTP reply for a destination.  At each hop, we used a
unique nonce \vsix source address, waiting up to three seconds for the NTP
reply.  Each nonce is 64 bits, chosen at random, and used to populate
the lower 64 bits of the source address.  
For subsequent correlation of 
back-scan activity, we recorded 
all $<$nonce, target, hop limit$>$ tuples. 
Other than modifying the
IPv6 layer headers (source address and hop limit), our NTP queries are 
standard and fully compliant with the specification.  

Between February 2025 and April 2026, we sent increasing Hop Limit NTP
requests 
to each of
these servers between 90 and 92 times, or approximately 6 times per month. Our NTP
request probes with increasing Hop Limits were sent at a rate of about 6,000
probes per day, well under 1 packet per second.
Concurrently, we ran \texttt{tcpdump} on our
server's Ethernet interface to detect incoming scans for any of the nonce
addresses that we had recorded. \texttt{tcpdump} reported that no packets were
dropped during our study.

Note that, given the approximately $d=2.45M$ probes sent with random
nonces, and a nonce space of $n=2^{64}\simeq1.84\e{19}$, the
approximate probability
of a nonce collision is vanishingly small:
$$
p \approx 1 - \exp\!\left(-\frac{d(d-1)}{2n}\right) \approx 1 - \exp\!\left(-\frac{(2.45 \times 10^6)^2}{2 \cdot 2^{64}}\right)
\approx 1.63 \times 10^{-7}
$$
\noindent Hence, we do not expect any false positives as a result of a 
collision between random nonces in our experiments. After each of our
experiments, including a 14-month study in \S\ref{sec:results} and a weeklong
evaluation in \S\ref{sec:casestudy}, we verified that no \vsix nonce address was
used more than once.

\begin{figure}[t]
    \centering
    \includegraphics[width=\linewidth]{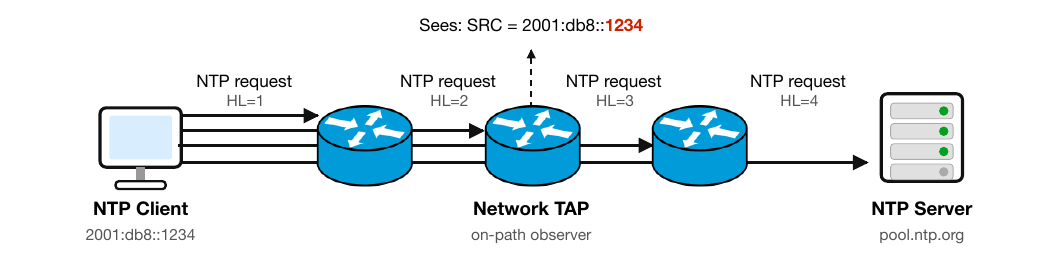}
    \caption{An on-path monitor detects traffic in the core of the network. This
    type of monitoring is restricted to \eg ISPs and law enforcement.}
    \label{fig:on-path-monitoring}
\end{figure}

\subsubsection{Detecting On-Path \vsix Back-Scanning}
\label{sec:on-path}

The traceroute-like approach to sending NTP requests that we implement is based
on a technique by Roberts and Plonka~\cite{roberts2020watching}. Roberts and
Plonka instrumented the \yarrpsix \vsix topology discovery tool to use a
different \vsix nonce address for each $<$Hop Limit, destination IP$>$ tuple.
Then, they monitored for a variety of network signals (\eg rDNS, active scanning,
etc.) in order to identify where on-path network surveillance occurs. We borrow
their technique of incrementing Hop Limits with different nonce addresses in
order to \emph{exclude} any back-scans in which an \emph{intermediate} hop --
that is, one that does not reach and receive a response from an NTP server -- is
targeted. Figure~\ref{fig:on-path-monitoring} depicts an on-path network monitor
that learns nonce addresses from traffic that passes through its routers.

On-path monitors are necessarily higher-powered than the type of monitor we
attempt to detect -- they must have access to core Internet infrastructure from
which to harvest \vsix addresses. This type of access is typically restricted to
powerful entities, such as nation-state intelligence or law-enforcement
agencies, or to \acp{ISP} or \acp{CDN} that own and operate the network
infrastructure.

\subsubsection{NTP Pool-Based \vsix Back-Scanning}

\begin{figure}[t]
    \centering
    \includegraphics[width=\linewidth]{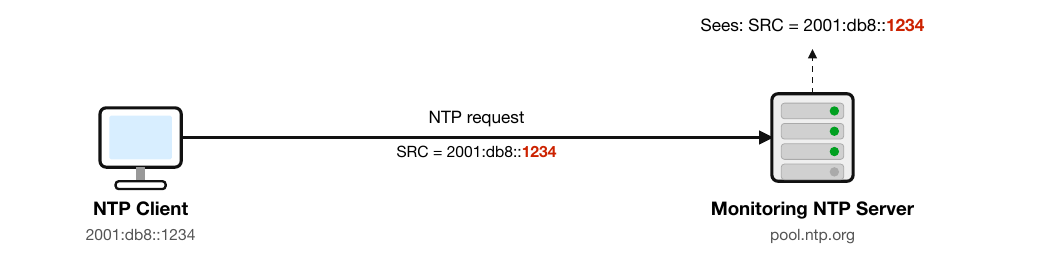}
    \caption{By contrast, an NTP Pool-based monitor need only operate an NTP
    server as part of the NTP Pool.}
    \label{fig:ntp-endpoint-monitoring}
\end{figure}

By contrast, we are interested in excluding these types of monitors and focus
instead on monitors who need only meet the low bar of joining the NTP Pool.
Prior work has shown that joining the NTP Pool can result in learning billions
of unique \vsix addresses over a period of only several months, making it a
natural means of \vsix reconnaissance to feed future active
scans~\cite{rye2023ipv6}. As opposed to the on-path network monitor in
Figure~\ref{fig:on-path-monitoring}, the NTP server endpoint network monitor we are
attempting to discover is represented in
Figure~\ref{fig:ntp-endpoint-monitoring}.

\begin{table*}[t]
  \centering
  \caption{NTP back-scanning clusters (endpoint observations only). \emph{Targets} = distinct NTP servers in cluster; \emph{Scanners} = distinct back-scanner IPs; \emph{Infra} = whether back-scanning came from the NTP server itself (Self), a separate third-party host (Third-party), or both (Mixed); \emph{Nonces} = distinct nonce addresses scanned; \emph{Packets} = total packets to endpoint-hop nonces.}
  \label{tab:clusters}
  \begin{tabularx}{\linewidth}{rrrrrr>{\raggedright\arraybackslash}X>{\raggedright\arraybackslash}X>{\raggedright\arraybackslash}X}
    \toprule
    \textbf{Cluster} & \textbf{Targets} & \textbf{Scanners} & \textbf{Infra} & \textbf{Nonces} & \textbf{Packets} & \textbf{Protocols} & \textbf{TCP ports} & \textbf{UDP ports} \\
    \midrule
    1 & 12 & 13 & Mixed & 204 & 481 & ICMPv6 & --- & --- \\
    2 & 8 & 3,198 & Third-party & 717 & 542,597 & TCP, UDP & 572~ports & 9~ports \\
    3 & 1 & 1 & Self only & 23 & 23 & ICMPv6 & --- & --- \\
    4 & 1 & 1 & Third-party & 1 & 2 & TCP & 5555~(adb) & --- \\
    \midrule
    \textbf{Total} & 22 & 3,213 & --- & 945 & 543,103 & --- & 572~ports & 9~ports\\
    \bottomrule
  \end{tabularx}
\end{table*}

To identify NTP servers that are using the induced traffic from the NTP Pool to
collect active \vsix addresses, we passively capture inbound traffic to any of
the nonce addresses that we used in our iterative Hop Limit NTP request
scanning. 

When we receive scans toward any of our nonce addresses, we first
check whether that nonce \vsix address reached an NTP server and received an NTP
reply. This step helps eliminate on-path back-scanners (\S\ref{sec:on-path}) and
avoids misattributing scanning behavior to NTP Pool server operators that are
not actively back-scanning themselves. For scans sent to \vsix nonce addresses
that did reach an NTP Pool server, we link the nonce IP address with the NTP
Pool server it queried and the source \vsix address that sent the scan traffic. 

We note that the scanning infrastructure may be entirely separate from the NTP
infrastructure used to learn \vsix addresses. As we will see in
\S\ref{sec:results}, they frequently are disjoint.  Furthermore, NTP
Pool servers operated by the same entity that are used to harvest active \vsix
addresses can be linked if they use the same scanning infrastructure; we consider
these NTP servers to be part of the same back-scanning ``cluster''.

\subsubsection{Back-Scanner Behavioral and Intent Analysis}

Finally, we conduct a limited experiment to better understand the behavior and
intent of NTP Pool-based back-scanners. For a weeklong period in April 2026, we
capture the TCP three-way handshake and initial data packet from NTP server
back-scanners. We use the data contained in the first data packet to understand
the application-layer behavior of NTP endpoint scanners, including what types of
exposed services and vulnerabilities they are attempting to detect.

\subsection{Limitations}
\label{sec:limitations}

There are several limitations of our methodology that we acknowledge may
affect the results we discuss in \S\ref{sec:results} and \S\ref{sec:casestudy}.

\textbf{Probe Rate:} First, the rate at which we send nonce NTP probes is
deliberately slow in order to avoid overburdening any of the NTP servers we query.
However, we note that some servers may sample from the \vsix addresses that
reach them, and only back-scan a subset of the unique \vsix addresses they
observe. If a back-scanner's sampling rate is low, it is possible that the
90--92 nonce addresses that we sent to each were insufficient to trigger
back-scanning.

\textbf{NTP Pool Server Completeness:} Second, we use the methodology employed
by several pieces of prior work to discover NTP Pool servers through repeated
DNS
queries~\cite{moura2024deep,jeitner2020impact,huang2025measuring,rytilahti2018masters}.
A major limitation of this strategy is that NTP Pool servers that choose to
receive smaller proportions of NTP traffic for their country ``zone'' are less
likely to be returned in DNS queries. NTP Pool servers that are in ``monitor''
mode or have been unreachable or unreliable will also not be returned in NTP
Pool DNS responses at all. If these kinds of servers are monitoring \vsix
traffic to initiate back-scans, we are less likely to discover them through
repeated DNS queries.

Recent work by Beverly and Rye~\cite{beverly2026ntpfool} demonstrated how to
obtain a complete snapshot of the NTP Pool via its API. We leave it to future
work to probe NTP Pool servers that we did not gather during our DNS
enumeration.  However, we note that the NTP back-scanning we present in
\S\ref{sec:results} and \S\ref{sec:casestudy} is a lower bound on the number of
back-scanning servers during the same time interval.

\textbf{Vantage Point Geographic Distribution:} A final limitation of our
methodology is our restriction to a single US cloud vantage point during our
experiments in \S\ref{sec:results} and \S\ref{sec:casestudy}. It is possible,
for instance, that an NTP Pool \vsix monitor chooses to back-scan only \vsix
addresses located in a specific country, or from a certain type of network (\eg
residential or cellular). If this is the case, we will not be back-scanned
unless this entity filters for US or cloud infrastructure. We discuss our plans
to expand and diversify our \vsix monitoring infrastructure in
\S\ref{sec:future}.

\section{Results}
\label{sec:results}

In this section, we describe the results of our one-year NTP iterative probing
campaign, including detecting \vsix-monitoring NTP servers, discovering their
scanning infrastructure, and clustering NTP server operators. Then, we examine
application-layer data from a weeklong packet capture to understand the types of
services and applications that the back-scanning operators are attempting to
discover.

\subsection{Identifying NTP Pool \vsix Monitors}

From February 2025 to April 2026, we iteratively probed 2,335 responsive NTP
servers that were joined to the NTP Pool, using \vsix Hop Limits that increased
from 1 to 32 or until an NTP response was received (\S\ref{sec:ntpqueries}). Each
server was iteratively probed between 90 and 92 times over the course of our
experiment. This resulted in an effective probe rate of approximately 1.5 NTP
queries per week; a negligible load on servers that receive many orders of
magnitude more traffic by participating in the NTP Pool.

Concurrently, we captured all inbound \vsix traffic destined for one of our
previously used nonce addresses. During this longitudinal experiment, we
responded to \icmpvsix Echo Requests with \icmpvsix Echo Replies; however, we
did not respond to any TCP or UDP traffic.

\subsubsection{Eliminating On-Path Back-Scanners} 

At the conclusion of our scanning campaign, we first determined whether any of
the back-scanned \vsix addresses were from probes which did not reach the
targeted NTP server (and thus, could not have been collected by the server but
rather some on-path monitor). This type of monitor detection has been studied in
detail by Roberts and Plonka~\cite{roberts2020watching}, though they did not use
NTP as a scanning protocol. Some 230 nonce \vsix addresses that did
not reach their target NTP server were back-scanned, and we removed these from
consideration.

\subsubsection{Clustering Monitoring NTP Servers} 

After removing on-path back-scans, 945 nonce \vsix addresses remained that had
been back-scanned after reaching an NTP Pool server. To understand NTP Pool
\vsix monitors, we linked the nonce \vsix address that we used with the NTP Pool
server \vsix address it reached. In most cases, the \vsix address that initiated
the back-scan was different from the NTP server that was queried, as well. By
grouping the NTP servers that were queried from our measurement platform with
the \vsix addresses used to initiate scans, we uncovered four distinct clusters
of NTP Pool monitors. These clusters are summarized in Table~\ref{tab:clusters}.

Cluster 1 was the largest cluster by number of participating NTP Pool servers,
with 12 distinct \vsix NTP servers. This cluster probed 204 different nonce
\vsix addresses and initiated only \icmpvsix Echo Requests. Cluster 1 used 13
different scanning \vsix addresses to initiate the back-scans; 12 of the 13 were
different from the NTP server addresses, while one of the NTP servers was itself
used to send \icmpvsix Echo Requests. This was the only cluster that exhibited a
``mixed'' back-scanning infrastructure in which both NTP servers and third-party
\vsix addresses were used to conduct the scanning.

Cluster 2 was the largest cluster by number of nonces back-scanned, number of
packets sent, and size of its back-scanning infrastructure. Cluster 2
comprised eight unique \vsix NTP Pool servers; these servers were used to
collect and back-scan 717 distinct nonce addresses sent from our measurement
platform. As opposed to Cluster 1, Cluster 2 conducted only TCP and UDP scans,
with 572 distinct TCP ports and nine distinct UDP ports scanned.  Nearly 3,200
unique source \vsix addresses initiated Cluster 2 back-scans, the largest set of
scanning infrastructure of any of the back-scanning clusters. Cluster 2 used
only infrastructure from Linode and DigitalOcean for its back-scanning; its NTP
servers were located in AWS IP space.

Clusters 3 and 4 each comprised only a single NTP Pool server. Cluster 3
initiated \icmpvsix Echo Requests for 23 distinct nonce addresses from the NTP
server itself. Cluster 4 initiated only two TCP SYNs to TCP port 5555
(traditionally Android Debug Bridge (ADB)) for a single \vsix nonce address
using third-party scanning infrastructure.

Table~\ref{tab:cluster_ntp_asns} lists the \acp{AS} and countries that those
\acp{AS} are associated with, as obtained from Team Cymru's Whois lookup
service~\cite{cymru}. The NTP servers used to collect active \vsix addresses are
all located in three cloud hosting providers: Amazon Web Services (AWS), Alibaba
Cloud, and Tencent Cloud.

Because almost all back-scans are sent from IP addresses that are disjoint from
the back-scanners' NTP servers, we also examine the \acp{AS} and Whois countries
of the back-scanning infrastructure. Table~\ref{tab:cluster_scanner_asns}
summarizes the back-scanning infrastructure details. While Tencent, Alibaba, and
AWS are all used for back-scanning, DigitalOcean and Linode are the primary
ASes used by Cluster 2 for its back-scanning. Cluster 4 is the only cluster to
use FranTech Solutions, another cloud hosting provider.

\begin{table}[t]
\caption{NTP servers per cluster by ASN and country.}
\label{tab:cluster_ntp_asns}
\centering\small
\begin{tabularx}{\linewidth}{c l X l r}
\toprule
    \textbf{Cluster} & \textbf{ASN} & \textbf{Organization} & \textbf{CC} &
    \textbf{\#IPs} \\
\midrule
  1 & AS45090 & TENCENT-NET-AP Shenzhen Tencent Computer Systems Company Limited & CN & 4 \\
   & AS45102 & ALIBABA-CN-NET Alibaba US Technology Co. & SG & 4 \\
   & AS16509 & Amazon.com & US & 2 \\
   & AS37963 & ALIBABA-CN-NET Hangzhou Alibaba Advertising Co. & CN & 1 \\
   & AS16509 & Amazon.com & IE & 1 \\
\addlinespace[3pt]
  2 & AS16509 & Amazon.com & US & 6 \\
   & AS16509 & Amazon.com & IE & 2 \\
\addlinespace[3pt]
  3 & AS45102 & ALIBABA-CN-NET Alibaba US Technology Co. & SG & 1 \\
\addlinespace[3pt]
  4 & AS45090 & TENCENT-NET-AP Shenzhen Tencent Computer Systems Company Limited & CN & 1 \\
\addlinespace[3pt]
\bottomrule
\end{tabularx}
\end{table}

\begin{table}[t]
\caption{Scanner IPs per cluster by ASN and country.}
\label{tab:cluster_scanner_asns}
\centering\small
\begin{tabularx}{\linewidth}{c l X l r}
\toprule
    \textbf{Cluster} & \textbf{ASN} & \textbf{Organization} & \textbf{CC} &
    \textbf{\#IPs} \\
\midrule
  1 & AS45090 & TENCENT-NET-AP Shenzhen Tencent Computer Systems Company Limited & CN & 4 \\
   & AS45102 & ALIBABA-CN-NET Alibaba US Technology Co. & SG & 4 \\
   & AS16509 & Amazon.com & US & 2 \\
   & AS37963 & ALIBABA-CN-NET Hangzhou Alibaba Advertising Co. & CN & 1 \\
   & AS16509 & Amazon.com & IE & 1 \\
   & AS23910 & CNGI-CERNET2-AS-AP China Next Generation Internet CERNET2 & CN & 1 \\
\addlinespace[3pt]
  2 & AS14061 & DigitalOcean & US & 1,548 \\
   & AS63949 & AKAMAI-LINODE-AP Akamai Connected Cloud & US & 1,428 \\
   & AS16509 & Amazon.com & US & 110 \\
   & AS14618 & Amazon.com & US & 108 \\
   & AS14061 & DigitalOcean & SG & 2 \\
   & AS63949 & AKAMAI-LINODE-AP Akamai Connected Cloud & SG & 2 \\
\addlinespace[3pt]
  3 & AS45102 & ALIBABA-CN-NET Alibaba US Technology Co. & SG & 1 \\
\addlinespace[3pt]
  4 & AS53667 & FranTech Solutions & US & 1 \\
\addlinespace[3pt]
\bottomrule
\end{tabularx}
\end{table}

\subsection{Characterizing NTP Pool Monitors}

Next, we characterize the behavior of NTP Pool-based \vsix monitors across
several axes.

\subsubsection{NTP Pool Metadata}
We examine metadata about the monitoring NTP servers available from
the \pool.  The \pool allows registration either with a public
username and email, or anonymously.  For all of the monitoring NTP
servers, both names are anonymous.  Continentally, 11 servers are
within the ``asia'' zone, 7 in ``north-america'', and 4 in ``europe.''
The most popular zone countries are: ``cn'' (6), ``us'' (5), ``uk''
(3), and ``sg'' (3).  Of interest, the median lifetime (length of time
the server has been registered within the \pool) is 484 days --
indicating that these servers (and their monitoring) have likely been
active and previously unnoticed for an extended period.

\subsubsection{Delay Before Back-Scanning}

A first feature of NTP Pool-based back-scanning entities is how much time
elapses between when the nonce address reaches the server and when the first
back-scan is initiated. To understand how quickly this occurs, we computed the
time elapsed between these two events for each of the 945 back-scanned nonce
\vsix addresses and grouped them by back-scanning cluster.

\begin{figure}[t]
    \centering
    \includegraphics[width=\linewidth]{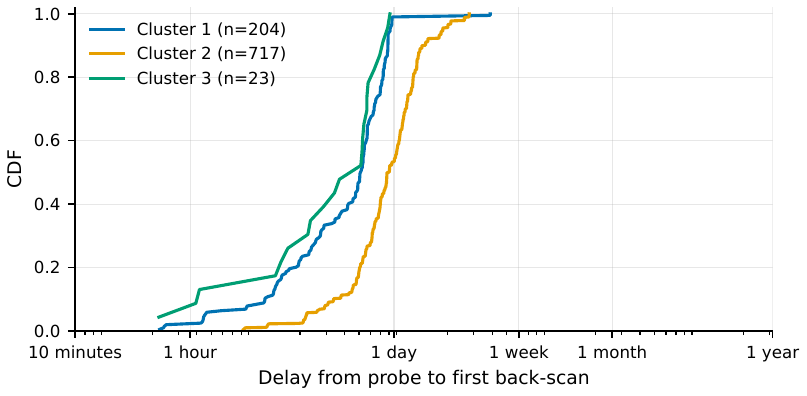}
    \caption{CDF of the delay between using a nonce address and when the nonce
    address was back-scanned by type of back-scanner.}
    \label{fig:delay_cdf}
\end{figure}

Figure~\ref{fig:delay_cdf} depicts CDFs of the time to first back-scan packet
for the 945 back-scanned nonces grouped by cluster. The median time before
back-scan is approximately 14 hours for both Clusters 1 and 3. Cluster 2's
median time to first back-scan was nearly twice as long, at 22 hours.
Approximately 10\% of Cluster 2 nonce addresses were not scanned back for a day
or more, as the long tail of the distribution indicates. Cluster 4 only scanned
back a single address after approximately 1 minute and is not displayed in
Figure~\ref{fig:delay_cdf}.

\begin{figure}[t]
    \centering
    \includegraphics[width=\linewidth]{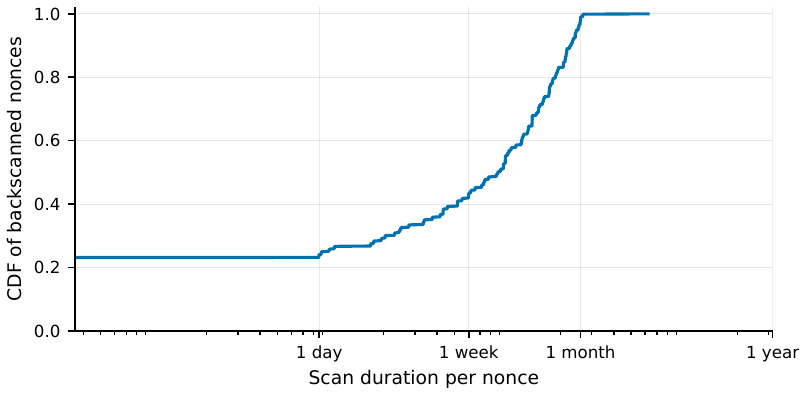}
    \caption{Scan duration by back-scanned nonce. 217/945 (23\%) nonces were
    back-scanned only once.}
    \label{fig:duration-cdf}
\end{figure}

\subsubsection{Duration of Back-Scanning}

Next, we examined how long a back-scanner would continue to scan the same nonce
address after it had scanned it initially -- what we refer to as the \emph{scan
duration}. We find that most of the scanning clusters scan a nonce address only
once. Only 10 of the 204 nonce addresses back-scanned by Cluster 1 received more
than a single packet; the other 194 had an effective scan duration of 0. All 23
nonce addresses back-scanned by Cluster 3 received only a single packet. Cluster
4, which scanned back only a single address, did send two TCP packets, but did
so within the same second.

Nonce addresses back-scanned by Cluster 2, however, displayed considerable scan
durations. Figure~\ref{fig:duration-cdf} plots the scan durations of the 945
nonces back-scanned by all four clusters. Cluster 2's back-scans comprise the
vast majority of the nonzero durations; the median scan duration is quite long,
at more than a week, and some back-scanning continues for up to a month.
Because we send only a single NTP request packet from each nonce address and do
not respond to TCP or UDP scans, this indicates that some
back-scanners have an intensive battery of back-scanning that they send to every
\vsix address selected for scanning over the course of one or more weeks. 

\subsubsection{Back-Scanner Scan Rate}

The next aspect of NTP endpoint back-scanner behavior we investigated was the
fraction of our nonce addresses that were back-scanned. Our intuition was that
some NTP back-scanners may sample and scan back only a small fraction of the
\vsix addresses they observe, either to evade detection by observers watching
for back-scans, or simply due to the inability to scan back all of the
source addresses they observe within a reasonable period of time.

\begin{figure}[t]
    \centering
    \includegraphics[width=\linewidth]{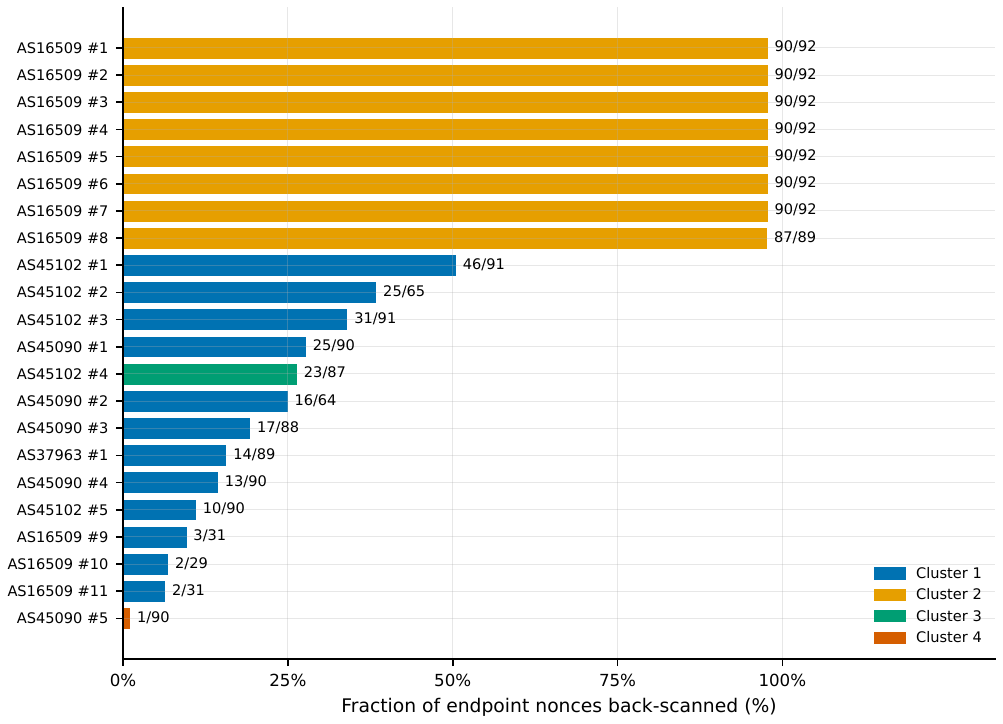}
    \caption{Fraction of nonce NTP requests that reached an NTP Pool server that
    generated back-scans, aggregated by NTP server target.}
    \label{fig:endpoint-target}
\end{figure}

Figure~\ref{fig:endpoint-target} displays the fraction of NTP nonces that
reached each of the 22 NTP servers that were back-scanned. Each cluster is identified by a
different color, and each NTP server is identified only by the \ac{AS} that its
address is from. Note that some back-scan servers have replied to fewer NTP
requests than others, which changes the denominator for their back-scan rate.

Cluster 2 is notable for a nearly 100\% back-scan rate; only two of the nonce
addresses that reached each of the Cluster 2 NTP servers were not back-scanned. 
By contrast, all of the other clusters exhibited substantially lower back-scan
rates. Only one other NTP back-scan server has a back-scan rate of more than
50\%; Cluster 4 is notable for having the lowest back-scan rate, with only one
of the 90 NTP requests that reached it ever receiving a back-scan.

This demonstrates considerable variety in back-scanning rates.
While Cluster 2 scanned nearly every nonce address it received from our server,
other clusters are more conservative in their approach to scanning back NTP Pool
clients.

\subsubsection{Back-Scanner Port Selection} 

Finally, for Cluster 2, we examine the set of destination TCP and UDP ports that
were probed (we exclude Cluster 4, which probed only one nonce address on TCP
port 5555). Cluster 2 was unique in both the number of nonce addresses probed
and the variety of transport-layer ports it sent traffic to -- over the 14
months of our experiment, Cluster 2 probed nonces on 572 TCP and 9 UDP ports.

\begin{figure}[t]
    \centering
    \includegraphics[width=\linewidth]{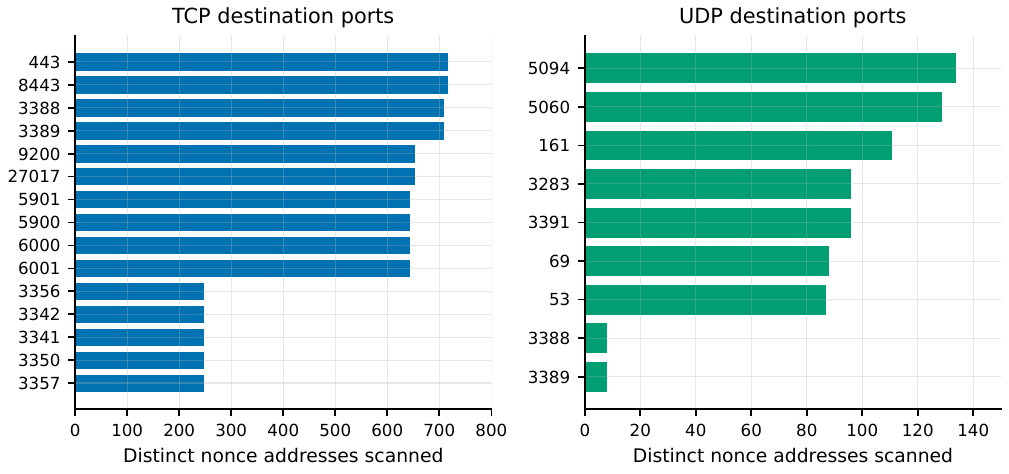}
    \caption{Top TCP and UDP ports back-scanned by number of unique nonce
    addresses targeted.}
    \label{fig:top_ports}
\end{figure}

Figure~\ref{fig:top_ports} depicts the number of nonces that were probed 
for the top 15 TCP ports and all 9 UDP ports. All 717 nonces that were scanned
by Cluster 2 were scanned on ports TCP/443 (HTTPS) and TCP/8443 (HTTPS-alt). This
perhaps indicates that Cluster 2 prioritizes capturing web content, given its
potential to identify the type of device that is responding to the scan. Ports
TCP/3388 and TCP/3389 (RDP) were targeted on 709 (99\%) nonce addresses, while
653 (91\%) nonce addresses had ports TCP/9200 (Elasticsearch) and TCP/27017
(MongoDB) scanned.

Only 9 UDP ports were scanned by Cluster 2, and the number of nonces that were
scanned across all of these ports was significantly smaller than in the TCP
scans. The most commonly scanned UDP port was UDP/5094 with 134 (19\%) nonce
addresses scanned. UDP/5094 is the well-known port for the Highway Addressable
Remote Transducer over IP (HART-IP) protocol, used for industrial automation.
UDP/5060 (SIP) was the next most common, with 129 (18\%) of nonce addresses
scanned. UDP ports 161 (SNMP), 3283 (Apple RDP), 3391 (RDP over UDP), 69
(TFTP), 53 (DNS), 3388 and 3389 (RDP) were all scanned on only 15\% of all nonce
addresses or less.

In general, we note that Cluster 2 focuses on ports that are either very common
for all device types (\eg HTTPS) or almost exclusively run by endpoint systems,
such as MongoDB, Elasticsearch, RDP, and industrial automation protocols like
HART.

\subsubsection{Back-Scanner Port Scanning Strategy}

\begin{figure}[t]
    \centering
    \includegraphics[width=\linewidth]{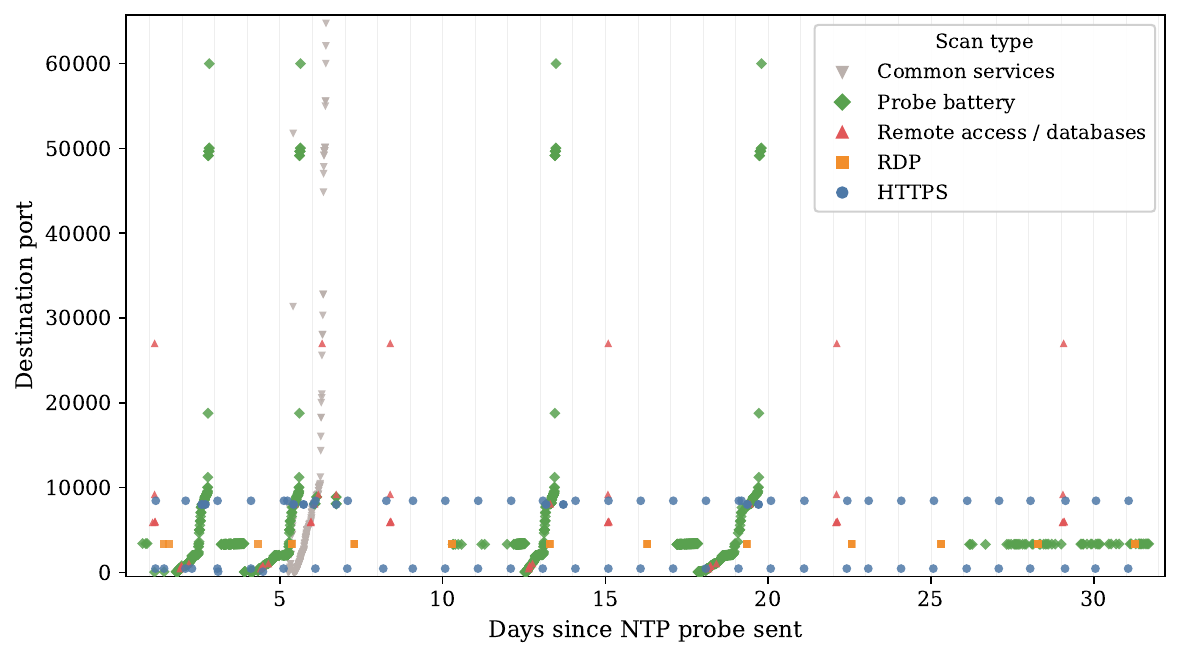}
    \caption{Back-scan timeline for a single NTP nonce address over 31 days.
    Each point represents one scan. Shape and color indicate scan type: HTTPS
    ports (443, 8443) are swept approximately daily; RDP ports (3388, 3389)
    roughly three days; remote access and database ports approximately weekly; a
    larger probe battery occurs four times within the first three weeks; and a
    thorough, common services enumeration only once.}
    \label{fig:single_nonce}
\end{figure}

Cluster 2 back-scans NTP nonce addresses according to a structured,
multi-tier schedule that is immediately visible from a single nonce's scan
timeline. Figure~\ref{fig:single_nonce} shows the back-scan history for a
single representative nonce over the 31 days following the NTP probe (other
nonce \vsix address back-scanning patterns are similar). Five
distinct scan types are apparent. HTTPS ports (443 and 8443) are revisited
approximately daily, reflecting continuous monitoring of TLS-enabled services.
RDP ports (3388 and 3389) recur every three days. A broader pass over remote
access and database services -- VNC, X11, and common database ports -- occurs
roughly weekly. The dominant visual feature is the \emph{probe battery}: a
fixed list of approximately 300 ports swept at five-minute intervals per port.
This probe battery occurs twice within the first week, and four times within
three weeks following the NTP probe. Finally, a single pass of a thorough,
common services enumeration sweep occurs once during the observation window.
Because no individual nonce is observed long enough to capture two such
passes, we cannot estimate this tier's inter-campaign interval from a single
nonce; cluster-level analysis across all nonces suggests it occurs infrequently
and on the order of one to four months apart.

\section{Application-Layer Back-Scan Analysis}
\label{sec:casestudy}

After completing our yearlong experiment focused on \emph{identifying} NTP
Pool-based network monitoring, we designed an additional short-term experiment
focused on understanding the intent behind back-scanning.  To do this, we
analyzed additional scan features beyond the port and transport-layer protocol
used by NTP Pool-based endpoint back-scanners, including the first data packet
captured after the TCP three-way handshake. 

\subsection{Data Collection}

The data for this experiment was captured during a weeklong period in April 2026
using a different /64, procured from a cloud hosting provider, than the one used
in \S\ref{sec:results}. Rather than logging only the initial TCP SYN sent by a
back-scanner, we further completed the three-way handshake with any inbound TCP
connection to one of our nonce \vsix addresses and captured the initial data
packet as well. This allows us to characterize the application-layer scanning
that a back-scanner is doing; for instance, while TCP SYNs to port 80 might be
expected to check for an HTTP server, a back-scanner may check for unexpected
protocols, as well.

During the week we conducted this case study, we received back-scans from 510
unique \vsix back-scanner infrastructure addresses directed to eight nonce
addresses. The nonces scanned were all received by one of the eight NTP servers
associated with Cluster 2 (\S\ref{sec:results}). All eight nonces that
were back-scanned received scans on 329 distinct ports.

\begin{table}[t]
\caption{Port-specific application-layer payloads sent by Cluster 2.  These
    payloads appear only on a single destination port, indicating deliberate
    service and manufacturer targeting.}
\label{tab:targeted_probes}
\centering\small
\begin{tabularx}{\linewidth}{r l X}
\toprule
    \textbf{Port} & \textbf{Service} & \textbf{Targeted probe payload} \\
\midrule
443   & HTTPS     & TLS ClientHello with unique random session IDs (per-connection) \\
8443  & HTTPS-alt & TLS ClientHello with unique random session IDs (per-connection) \\
9200  & Elastic   & HTTP \texttt{GET /\_nodes}, \texttt{/\_nodes/\_local}, \texttt{/\_stats} via \texttt{python-urllib3/2.6.3} \\
9443  & HTTPS-alt & HTTP \texttt{GET /sonicui/7/sslvpn-portal/} (SonicWall SSL-VPN probe) \\
27017/27018 & MongoDB   & MongoDB wire protocol \texttt{admin.\$cmd isMaster}; also TLS ClientHello \\
\bottomrule
\end{tabularx}
\end{table}

\begin{table}[h]
\caption{Cluster 2 probe battery application-layer payloads: each of 38 payloads
    is sent to every responsive TCP port within the probe battery set of 327 TCP
    ports.}
\label{tab:probe_battery}
\centering\small
\begin{tabular}{ll}
\toprule
    \textbf{Category} & \textbf{Probe} \\
\midrule
  \multicolumn{2}{l}{\textit{RDP}} \\
  & \texttt{RDP X.224 (mstshash=beio)} \\
  & \texttt{RDP X.224 bare CR} \\
  & \texttt{RDP X.224 (mstshash=Administrator)} \\
\addlinespace[3pt]
  \multicolumn{2}{l}{\textit{TLS}} \\
  & \texttt{TLS 1.0 ClientHello (v1)} \\
  & \texttt{TLS 1.0 ClientHello (v2)} \\
\addlinespace[3pt]
  \multicolumn{2}{l}{\textit{HTTP/Web/SIP}} \\
  & \texttt{HTTP GET /} \\
  & \texttt{HTTP OPTIONS /} \\
  & \texttt{RTSP OPTIONS /} \\
  & \texttt{HTTP Nmap probe} \\
  & \texttt{SIP OPTIONS} \\
\addlinespace[3pt]
  \multicolumn{2}{l}{\textit{Windows/AD}} \\
  & \texttt{Kerberos AS-REQ (krbtgt)} \\
  & \texttt{SMB1 Negotiate} \\
  & \texttt{MMS (Windows Media Services)} \\
  & \texttt{MS-SQL pre-login} \\
  & \texttt{Microsoft MSMQ (MS-MQQB)} \\
\addlinespace[3pt]
  \multicolumn{2}{l}{\textit{Database}} \\
  & \texttt{LDAP SearchRequest} \\
  & \texttt{LDAP Anonymous Bind Request} \\
  & \texttt{Java RMI} \\
  & \texttt{Oracle TNS Connect} \\
  & \texttt{CORBA IIOP} \\
\addlinespace[3pt]
  \multicolumn{2}{l}{\textit{Network}} \\
  & \texttt{Sun RPC portmapper} \\
  & \texttt{DNS CHAOS version.bind} \\
  & \texttt{DNS Status Request} \\
  & \texttt{AFP DSIGetStatus} \\
  & \texttt{OpenVPN TCP} \\
  & \texttt{X11 Display Connection Setup} \\
  & \texttt{NFS/RPC getport} \\
\addlinespace[3pt]
  \multicolumn{2}{l}{\textit{Specialty}} \\
  & \texttt{TNMP (Thomson/Trend Micro)} \\
  & \texttt{Tibco Rendezvous} \\
  & \texttt{Lotus Notes RPC} \\
  & \texttt{iSCSI Login Request} \\
  & \texttt{DICOM A-ASSOCIATE-RQ} \\
\addlinespace[3pt]
  \multicolumn{2}{l}{\textit{Generic}} \\
  & \texttt{Bare CRLF (banner grab)} \\
  & \texttt{HELP banner grab} \\
  & \texttt{'.default.' probe} \\
  & \texttt{Null probe} \\
  & \texttt{'.I20100.' probe} \\
\addlinespace[3pt]
  \multicolumn{2}{l}{\textit{Unknown}} \\
  & \texttt{Unknown binary (156B, +{<}M magic)} \\
\bottomrule
\end{tabular}
\end{table}

\subsection{Back-scan Payloads}

\subsubsection{Targeted Application-Layer Payloads}
\label{sec:targetedpayloads}

Several application-layer payloads were used on only a single TCP port, as
summarized in Table~\ref{tab:targeted_probes}. The TLS
ClientHellos sent to ports 443 (HTTPS) and 8443 (HTTPS-alt) used different
random session IDs each time they were sent, which occurred approximately daily
(Figure~\ref{fig:single_nonce}).
Other payloads, such as HTTP \texttt{GET} requests for the \texttt{/\_nodes},
\texttt{/\_nodes/\_local}, and \texttt{/\_stats} endpoints sent to TCP/9200,
provide introspection into Elasticsearch clusters -- \texttt{/\_nodes} returns
information about all nodes in an Elasticsearch cluster,
\texttt{/\_nodes/\_local} returns information specific to the node being
queried, and \texttt{/\_stats} returns statistics for the cluster, including
counts of documents and the store size. The HTTP \texttt{GET
/sonicui/7/sslvpn-portal/} on TCP/9443 indicates that Cluster 2 is actively
attempting to detect specific products, as this endpoint is unique to SonicWall
SSL-VPNs. Finally, Cluster 2 also attempts to run the MongoDB command
\texttt{admin.\$cmd isMaster}, which returns a MongoDB document containing
information about the node and its wider ``replica set'', if it is part of one.

These targeted application-layer payloads indicate that Cluster 2 is searching
for specific types of devices and services running on them, rather than simply
enumerating open ports.  

\subsubsection{Probe Battery Application-Layer Payloads}
\label{sec:genericpayloads}

In contrast with the targeted application-layer payloads we observed for a
handful of specific ports scanned by Cluster 2 (\S\ref{sec:targetedpayloads}),
their scans also indicate a more generic approach to service
discovery. During the ``probe battery'', in which a variety of ports are queried
in short succession (green vertical bars in Figure~\ref{fig:single_nonce}), each
responsive port receives a wide variety of application-layer payloads to attempt
to discover the service listening on that port. 

Each of 327 ports was scanned using each of the 38 payloads enumerated in 
Table~\ref{tab:probe_battery}, which organizes the payload types by category.
Table~\ref{tab:probe_battery_explained} in Appendix~\ref{sec:addldata} has
additional details about each of the payloads and what service they are designed
to elicit a response from, if known. The fact that each of these 38
application-layer payloads -- ranging from HTTP to database protocols to
Windows-specific applications -- is tried on each of 327 TCP ports indicates
that Cluster 2 is
attempting to find common applications, even when they are served over
nonstandard ports. 

The probe payloads used by Cluster 2 were tried in a fixed order, consistent
across all destination ports and target addresses. Table~\ref{tab:probe_order}
lists the 38 payloads in the order they were observed on the wire. Notably,
there are several interesting implementation differences between the targeted
payloads used in the scans from \S\ref{sec:targetedpayloads} and the payloads
delivered by the probe battery. For instance, the \texttt{ClientHello} frames
from TLS handshakes used during the daily scans each contain a unique session
ID. However, the TLS \texttt{ClientHellos} used in the probe battery's
payloads omit the field entirely. This suggests that the software used to
conduct the TLS scans in each of these probe types differs.

\begin{table}[t]
\caption{Observed transmission order Cluster 2's probe battery.}
\label{tab:probe_order}
\centering\small
\begin{tabular}{rll}
\toprule
    \textbf{\#} & \textbf{Category} & \textbf{Probe} \\
\midrule
   1 & Generic      & \texttt{Bare CRLF (banner grab)} \\
   2 & HTTP/Web/SIP & \texttt{HTTP GET /} \\
   3 & HTTP/Web/SIP & \texttt{HTTP OPTIONS /} \\
   4 & HTTP/Web/SIP & \texttt{RTSP OPTIONS /} \\
   5 & Network      & \texttt{Sun RPC portmapper} \\
   6 & Network      & \texttt{DNS CHAOS version.bind} \\
   7 & Network      & \texttt{DNS Status Request} \\
   8 & Generic      & \texttt{HELP banner grab} \\
   9 & TLS          & \texttt{TLS 1.0 ClientHello (v2)} \\
  10 & RDP          & \texttt{RDP X.224 (mstshash=beio)} \\
  11 & TLS          & \texttt{TLS 1.0 ClientHello (v1)} \\
  12 & Windows/AD   & \texttt{Kerberos AS-REQ (krbtgt)} \\
  13 & Windows/AD   & \texttt{SMB1 Negotiate} \\
  14 & Network      & \texttt{X11 Display Connection Setup} \\
  15 & HTTP/Web/SIP & \texttt{HTTP Nmap probe} \\
  16 & Generic      & \texttt{'.default.' probe} \\
  17 & Database     & \texttt{LDAP SearchRequest} \\
  18 & Database     & \texttt{LDAP Anonymous Bind Request} \\
  19 & HTTP/Web/SIP & \texttt{SIP OPTIONS} \\
  20 & Specialty    & \texttt{TNMP (Thomson/Trend Micro)} \\
  21 & RDP          & \texttt{RDP X.224 bare CR} \\
  22 & Specialty    & \texttt{Tibco Rendezvous} \\
  23 & Specialty    & \texttt{Lotus Notes RPC} \\
  24 & Database     & \texttt{Java RMI} \\
  25 & Windows/AD   & \texttt{MMS (Windows Media Services)} \\
  26 & Database     & \texttt{Oracle TNS Connect} \\
  27 & Windows/AD   & \texttt{MS-SQL pre-login} \\
  28 & Network      & \texttt{AFP DSIGetStatus} \\
  29 & Database     & \texttt{CORBA IIOP} \\
  30 & Network      & \texttt{OpenVPN TCP} \\
  31 & Windows/AD   & \texttt{Microsoft MSMQ (MS-MQQB)} \\
  32 & Unknown      & \texttt{Unknown binary (156B, +{<}M magic)} \\
  33 & Network      & \texttt{NFS/RPC getport} \\
  34 & Specialty    & \texttt{iSCSI Login Request} \\
  35 & Specialty    & \texttt{DICOM A-ASSOCIATE-RQ} \\
  36 & RDP          & \texttt{RDP X.224 (mstshash=Administrator)} \\
  37 & Generic      & \texttt{Null probe} \\
  38 & Generic      & \texttt{'.I20100.' probe} \\
\bottomrule
\end{tabular}
\end{table}

\section{Validation}
\label{sec:validation}

We identified the operator of the most prolific scanning cluster (Cluster 2, see
Table~\ref{tab:clusters}), comprising eight NTP Pool endpoint back-scanners
and over 3,000 back-scanning IPs, via PTR records associated with some of their
scanning infrastructure. Then, we contacted them via email to ask whether i)
they had communicated their use of the NTP Pool for back-scanning, ii) what they
used the back-scan data for, and iii) whether users could opt out of being port
scanned by their infrastructure.

This back-scanner was forthright that they were using the NTP Pool as a
mechanism to learn in-use \vsix addresses for back-scanning. They did not
directly address whether they had communicated their use of the NTP Pool with
the Pool maintainers; our conversations with the NTP Pool administrators
indicate that they did not. The scanner replied that they use the data collected
from port scans of NTP Pool clients for purely defensive purposes; \eg to
inform cybersecurity insurance risk calculations. 

Regarding the ability to opt out of their scans, we noted that detecting port
scans from this scanner is particularly difficult, due to three factors.  First,
Cluster 2 operates disjoint NTP infrastructure and back-scanning infrastructure
-- their scanning infrastructure is located in the IP space of the cloud
providers Linode and DigitalOcean, while their NTP infrastructure was hosted
by AWS.
Second, many of the \vsix addresses from which they back-scanned lacked rDNS
entries. Third, neither their back-scanning nor NTP infrastructure hosted a web
page to indicate their identity or how a user might opt out of their back-scans.

The operator agreed with our observation that a significant proportion of their
scanning infrastructure lacked rDNS entries. They indicated that $\sim$40\%
of their infrastructure was missing PTR records, which they claimed
they had subsequently corrected. Additional back-scan detection by our detection
infrastructure confirms that they did add rDNS entries for the majority of their
scanning infrastructure; their NTP Pool servers, however, are still not
identified using rDNS and neither the NTP servers nor scanning servers host a
webpage indicating who they are or how to opt out of their scans. They did
note, however, that the second-level domain in their rDNS entries redirects to a
page informing users about the nature of their scans and how to opt
out. 

\section{Recommendations}
\label{sec:recommendations}

The NTP Pool has been used to source \vsix addresses for back-scanning before;
in 2016, the NTP Pool administrators removed the network device search engine
Shodan's servers from the NTP Pool after they were observed conducting back-scans
to \vsix
addresses~\cite{shodanscanning,shodanseclists,shodangoodin,shodannull,shodanyc}
(none of the back-scanning clusters we detected is Shodan, to the best of our
knowledge). The fact that it has reappeared a decade later, via both a commercial
entity (\S\ref{sec:validation}) and academic research using the NTP Pool to port
scan active addresses~\cite{klopsch2025time}, coupled with \vsix's continued
growth, suggests that it is time to establish norms regarding the NTP Pool's use
for \vsix intelligence.

More generally, TCP and UDP port scanning of the type conducted by Clusters 2
and 4 elicits a variety of viewpoints. It is legally and ethically murky, as
courts have had wide-ranging interpretations of what constitutes ``access'' to a
computer system, and whether such access is
``authorized''~\cite{kerr2003cybercrime}.  In the US, port scanning
is frequently compared to ``knocking on doors'' to determine whether anyone
is home~\cite{kerr2016norms}, which is clearly not criminal behavior. However,
American citizens have had criminal charges filed and civil suits brought against
them for port scanning, though the charges were eventually
dropped~\cite{uscode1030,moulton}. In the UK, the 1990 Computer Misuse Act
covers unauthorized access offenses to computer systems~\cite{acts1990computer};
while in practice port scanning does not bring charges, legal analyses suggest
that even defensive organizations like \acp{CSIRT} may not be within the law to
test for server susceptibility to vulnerabilities like
Heartbleed~\cite{heartbleed} without explicit consent, due to detection
requiring the transmission of ``unusual
parameters''~\cite{cormack2014can,cormack2015internet}. Oh and Lee report that
port scanning is legally punishable in South Korea as the precursor to an
attack~\cite{oh2014need}.

In this section, we make several concrete recommendations for how to address the
rising phenomenon of NTP Pool-based \vsix back-scanning.

\subsection{Commercial and Research Entities}

Entities wishing to use the NTP Pool for \vsix address discovery for commercial
or research purposes -- for instance, to seed the types of security scans we
observed Cluster 2 initiating in this work -- should adhere to several
guidelines. First, they should disclose their use of the NTP Pool to the Pool
administrators. At minimum, the prospective Pool users should disclose the type
and duration of their experiments and data collection, the types of data they
are collecting, and what their intentions are with the data they collect. Entities
whose usage of the NTP Pool is not approved by the NTP Pool administrators
should respect the administrators' decision. Second, entities should make
themselves readily identifiable via PTR DNS records for all of their
infrastructure. While
the Cluster 2 operator identified itself on some of its scanning infrastructure,
it admitted that a significant fraction of its scanning addresses (40\%) did
not have rDNS entries (which it has since rectified), and its NTP Pool
servers are still not identified. Hosting web pages on NTP and scanning
infrastructure to explain who the scanning entity is, what the purpose of their
scanning is, and how to opt out, is another way in which ethical commercial and
research entities might participate in the NTP Pool.

NTP Pool clients -- many of whom are embedded or \ac{IoT} devices whose
operators are unaware that their devices query the NTP Pool to begin with -- did
not opt in to being port-scanned as a consequence of time synchronization.
Approved research that uses the NTP Pool to connect back to NTP clients should
avoid publishing the raw client \vsix addresses, as well as any other network
data that could be used to identify an individual or entity. Commercial products
that leverage data from re-connecting to NTP clients should be carefully
scrutinized to ensure they do not harm the users whose data they are collecting.
We also note that commercial entities that collect this data may have
obligations under privacy laws such as \ac{GDPR} and \ac{CCPA}.

\subsection{NTP Pool Administrators}
\label{sec:ntpadmins}

The NTP Pool administrators are in a unique position to reduce back-scanning. 
The NTP Pool uses a scoring system to determine whether servers that have joined
the Pool should continue to receive NTP client traffic by being included in DNS
responses to the NTP Pool's DNS infrastructure. Today, that scoring system
measures the availability (\ie unresponsive NTP servers are penalized) and
accuracy (\ie NTP servers providing bad time are penalized) of servers in the
Pool. However, evidence of back-scanning NTP clients could also be incorporated
into the scoring algorithm. The NTP Pool administrators have provided positive
feedback for this suggestion and we are actively working on integrating a
production version of our prototype from this paper into the NTP Pool scoring
system at the time of writing.

Additional measures the NTP Pool administrators might take include publishing an
\ac{AUP} that clarifies for would-be Pool operators what types of experiments or
data collection are considered acceptable. While unscrupulous actors might not
adhere to the \ac{AUP}, ethical research standards would preclude using the NTP
Pool without the Pool administrators' approval. A further, more administratively
laborious suggestion is that the NTP Pool might form a research review committee,
like the Tor Research Safety Board~\cite{torresearch}, that vets, approves, and
tracks ethical research proposals.

\subsection{Individual Users and Network Administrators}

Individual users and network administrators can take precautions to protect
themselves and their users from unwanted back-scanning sourced from NTP Pool
servers. While opting to not use the NTP Pool at all is a potential remediation,
the vast majority of NTP Pool servers did not initiate any back-scanning
during the course of our experiment.  We hope that the back-scanning
detection system we are integrating with the NTP Pool's scoring
system~(\S\ref{sec:ntpadmins}) will eliminate the need for users to actively filter
\vsix back-scanning NTP servers. 

However, in the interim, we will release a blocklist of NTP servers from which
we detect back-scanning on a periodic basis through our organization's website.
These blocklists will enable users to block traffic going to the NTP servers
that are learning \vsix addresses, rather than blocking the IP addresses
originating the scan traffic. As Cluster 2 demonstrates, the scanning
infrastructure employed by an NTP Pool-based back-scanner may be orders of
magnitude larger than the infrastructure collecting the addresses to back-scan.
This both makes it easier to block and prevents the back-scanning entity from
passively learning active addresses in the network.

\section{Conclusions and Future Work}
\label{sec:concl}

Invasive port scanning is routine in \vfour; the advent of network scanning
tools like \zmap makes exhaustive \vfour address space enumeration trivial.
In \vfour, the missions and business models of many companies (\eg
Shodan~\cite{shodan}, Censys~\cite{censys}, Shadowserver~\cite{shadowserver},
and others) revolve around exhaustive port, reachability, and vulnerability
scanning.

\vsix's massive address space precludes exhaustive scanning. Further, \vsix
clients tend to have short-lived and ephemeral addresses, complicating the
challenge of active address discovery. To adapt, researchers, industry, and
attackers alike need to develop strategies to \emph{learn} which addresses are
active before performing the same types of scans they perform routinely in
\vfour. 

In this work, we perform a longitudinal measurement of the NTP Pool to detect
NTP Pool-based \vsix monitoring over the course of 14 months. We detect four
distinct entities that use the NTP Pool as a mechanism to learn active \vsix
addresses (\S\ref{sec:results}). Each of the monitoring entities exhibits
significantly different monitoring and back-scanning behavior. For instance,
Cluster 1 is the largest by number of NTP servers used to collect \vsix
addresses, but initiates back-scans only using \icmpvsix Echo Requests. Cluster
2 also runs a large set of NTP Pool servers, and back-scans a wide variety of
TCP and UDP ports. Clusters 3 and 4 are single NTP server monitors, but scanned
back with \icmpvsix and TCP, respectively. We characterize these scanning
clusters across a variety of axes to understand from where, when, and how long
each of these back-scanners performs its probing. In \S\ref{sec:casestudy}, we
perform an in-depth analysis of the most prolific back-scanner (Cluster 2),
including investigating what \emph{applications} it scans for, using a weeklong
capture of the first data packet it sent to our reverse-monitoring
infrastructure.

In \S\ref{sec:validation}, we detail our discussion with the operator of
Cluster 2, who confirmed their operation of Cluster 2's infrastructure. They
additionally provided information about what they were using the data for, and
updated their scanning infrastructure's PTR records in response to our
inquiry.   

We described our efforts to improve the security and privacy of the NTP Pool, a
piece of critical Internet infrastructure, through our conversations and ongoing
work with the NTP Pool administrators. Our NTP Pool-based \vsix monitoring
detection system is in the early stages of being integrated with the NTP Pool's
scoring system, and will be an input into whether a volunteer's server will
continue to be a part of the NTP Pool. 

Finally, we close with some recommendations for entities that want to use the
NTP Pool for \vsix address discovery. We believe that the NTP Pool provides
unparalleled insight into the state of \vsix deployment, especially with respect
to \vsix client devices. Nonetheless, research utility and business advantage
must be balanced with ensuring the security and privacy of NTP Pool clients,
the majority of whom did not consciously decide to use the NTP Pool or know how
to pursue opting out of invasive back-scanning even when opt-outs are available.

We discuss the ethical considerations of this work in the Appendix. 

\subsection{Future Work}
\label{sec:future}

We are in the initial stages of integrating our NTP Pool \vsix monitoring
platform with the NTP Pool system. As part of this effort, we
are exploring other features that NTP Pool-based monitors might use to decide
whether to initiate back-scanning to an NTP client \vsix address.  

For instance, in this work we ran all of our experiments from a US-based cloud
hosting provider -- it is unclear whether geographically distributed servers
will detect country- or region-specific back-scanning behavior. Whether
monitoring detection run from non-cloud hosting IP space (\eg from a
residential monitor) might detect different back-scanning entities or
strategies is another avenue for future work we intend to pursue.  

Finally, while the NTP Pool presents a low barrier-to-entry way for prospective
monitors to learn active \vsix addresses, it is by no means the only one. We
intend to explore the use of other protocols and different crowdsourced services
to see whether they are also being used for \vsix monitoring.

\section*{Acknowledgments}
We thank Ask Bjørn Hansen and Steven Sommars for NTP Pool operational
support and insight.  This research was partially
supported by the NSF under grant OAC-2613546.
Views and conclusions contained in this document are those of the
authors and should not be interpreted as representing the
official policies, either expressed or implied, of NSF or the
U.S.\ government.

\bibliographystyle{ACM-Reference-Format}
\bibliography{conferences,refs}

\appendix %

\section{Open Science} %

To support the goal of open and reproducible science, we are releasing the
following artifacts to the community:

\begin{enumerate}
    \item The code necessary to generate nonced \vsix NTP queries on a per \vsix
        Hop Limit and NTP server combination basis.
    \item Scripts used to cluster \vsix NTP server scanners with their scanning
        infrastructure.
\end{enumerate}

These artifacts can be found at
\url{https://github.com/Network-Security-Privacy-Research/Cleaning-the-NTP-Pool-artifacts}.

Given the sensitivity of our detection infrastructure and the ephemeral nature
of \vsix addresses, we do not release the raw scan data from this study.

\section{Ethical Considerations} %

\subsection{Stakeholder Analysis and Risk Mitigation}

Our measurement methodology involves active probing of public NTP servers and
passive capture of unsolicited inbound traffic, and touches the interests of
several distinct stakeholder groups.

\textbf{NTP Pool Administrators and Server Operators:} The NTP Pool is a
volunteer-operated public infrastructure, and generating additional query load
on NTP Pool servers could impose a real cost to operators. We mitigated this by
rate-limiting our probes to one NTP request per \vsix Hop Limit every three
seconds, with each NTP server target scanned approximately once every ten days
-- a rate well under the normal range of NTP client behavior and negligible in
terms of an overall NTP Pool server's load. Our NTP request probes are
indistinguishable in form from ordinary client queries.  We notified the NTP
Pool project administrators of our study and supplied them with its results. One
of the primary contributions of this study is a continuous monitoring system
that tracks and provides alerts when NTP Pool servers are using the NTP Pool for
\vsix address back-scanning. 

\textbf{Back-Scanners:} The entities conducting back-scans are, by definition,
responding to our probe traffic in an unsolicited and arguably unexpected
manner. In the endpoint-observer case, these are NTP server operators or network
administrators with access to server logs; in the on-path case, they are network
operators conducting passive traffic monitoring. Our data collection is limited
to traffic arriving at addresses we control and that no legitimate service is
listening on.  We do not disclose the identities of individual back-scanners or
their precise network locations in this paper beyond the ASN level, consistent
with community norms for responsible disclosure of network measurement findings.
Raw IP addresses of back-scanners are retained only in the authors' private
dataset and are not published.

\textbf{NTP Pool Users:} Members of the public who rely on the NTP Pool for time
synchronization are not directly affected by our probing, as our queries are
processed identically to any other client request. The more significant concern
is whether our findings degrade the privacy of NTP Pool users generally or
reduce confidence in the NTP Pool -- for instance, if back-scanning behavior is
widespread enough that client addresses are being systematically harvested by
third parties. Our results suggest this risk is real for users who reach certain
NTP servers. To support user privacy, we will publish network ranges from which
we receive back-scans in order to support blocklisting. We are also working with
the NTP Pool administrators to flag and reduce or eliminate client traffic to
NTP Pool-sourced \vsix address harvesting that is later used in back-scanning.

\textbf{Data Handling:} All packet captures were stored on machines accessible
only to the research team. Captures contain traffic to addresses we controlled
and do not include user traffic or third-party communications. Raw IP addresses
appearing in back-scan data are used internally for analysis but are aggregated
to the ASN level for all published results. We will make the NTP scanning and
analysis code available for reproducibility; raw captures will not be published.

\section{Stateless Nonces}

Since the collection of the data in this paper, our nonce creation and
evaluation have evolved to be more efficient.  Specifically, the system
now implements \emph{stateless} nonces.  Recall that the purpose of
the nonce is to correlate inbound scans to the original NTP client
request that induced the scan, including the time of the request and
to which NTP server it was sent.  

Let $ts$ be a 5B field with the number of seconds since the UNIX
epoch, $ttl$ be a 1B field with the originating IPv6 hop limit, and
$sid$ be a 2B field with the \pool server identifier.  We form the
following 8 bytes by concatenating these three fields:
$$
   s = ts | ttl | sid
$$

We maintain an 8B secret key, $k$.  We use the 64-bit DES block cipher
and denote $E_k$ and $E^{-1}_k$ as the DES encryption and decryption
using key $k$, respectively.  Because DES is a block cipher, it
provides a keyed random permutation from an 8B input space to an 8B
output space.  Note that since we are only encrypting a single block,
there is no need for a block cipher mode.  Let $|$ denote
concatenation.  Instead of maintaining the nonces in a database, we
create the nonce when issuing the NTP query as:
$$
  nonce(ts,ttl,sid) = E_k(s)
$$

For each incoming packet, given the secret key, we can trivially
decode the nonce (the lower 64 bits of the IPv6 destination address)
to recover the original NTP query's timestamp, hop limit, and
destination:
$$
  ts,ttl,sid = E^{-1}_k(nonce)
$$

\section{NTP Server Operator Perspectives}

In our discussions with the primary developers who operate the 
core \pool infrastructure (\eg the database of participating servers,
time accuracy monitoring, and the DNS serving \texttt{ntp.org} that
maps client NTP requests to participating \pool servers), they 
encouraged us to share our findings with \pool members running
NTP servers.  We shared a summary of our technical approach
and findings to the \pool discourse message board
(\url{https://community.ntppool.org/}).  In addition, we 
provided a survey to collect opinions and perspectives from
the broader community operating NTP servers.  The survey collected
no personally identifiable information.

We received 13 responses, 12 of which indicated that they 
operate a server in the \pool.  Eight of 13 respondents (62\%) 
indicated that IPv6 addresses of NTP clients are more privacy
sensitive than their IPv4 addresses.  Interestingly, 
opinions varied as to whether back-scanning of IPv6 NTP
clients is malicious, and the response also varied depending
on the back-scanning protocol.  For instance, while eight of
13 indicated that ICMPv6 is malicious, 12 of 13 found 
TCP SYN back-scanning malicious.  

The variety of perspectives was further evident in the 
forum discussion.  While several operators expressed 
surprise and concern, e.g., ``Using the infrastructure to harvest
client IPs and execute active scans, even for connectivity diagnostics
or mapping purposes, constitutes an unsolicited and potentially
malicious interaction. I consider this an unacceptable use,''  
others argued that the endpoints themselves should be better
protected: ``What is the problem if they do? If your systems are
secure, you have nothing to worry about. Enlighten me please, why
scanning is a problem?''  In this same vein, a third operator 
commented: ``Any decent customer grade router is blocking all inbound
traffic by default, so knowing the client IP address does not help the
attacker.''  

These responses not only provide a variety of opinions, but also 
highlight unique properties and realities of current IPv6 deployments.
In particular,
NTP clients are frequently unwitting users of the \pool and are not necessarily
able to protect themselves. Prime examples, as we allude to in
this work, are IoT devices, e.g., thermostats, lightbulbs, cameras,
printers, etc., that are running embedded Linux and configured to use
the pool.  As related work~\cite{firewalls2026} has shown, many
residential networks use CPE that do not provide any
stateful firewall for IPv6. As the IPv6 addresses within the home are
public and globally routable, they are externally reachable.  Thus,
adversaries have a strong motivation to back-scan: if the random,
privacy extension client address can be discovered, it is frequently
possible to directly connect to the device, enumerate its
capabilities, or even exploit it. 

Our final survey question asked whether the \pool should adopt an
Acceptable Use Policy that bans servers that enable IPv6 back-scanning
of NTP clients.  All 13 respondents (100\%) indicated
``yes.''  While these respondents may be self-selecting and not
representative of the entire population of participating 
NTP server operators, our findings suggest that there is support
for the \pool to adopt stricter measures to prevent it from
being used as a vector for IPv6 back-scanning.

\section{Additional data}
\label{sec:addldata}

\begin{table*}[t]
\caption{Cluster 2's probe battery: all 38 payloads are sent to every
open TCP port discovered by a prior SYN scan, regardless of expected service.}
\label{tab:probe_battery_explained}
\centering\footnotesize
\setlength{\tabcolsep}{4pt}
\begin{tabularx}{\textwidth}{l l X}
\toprule
    \textbf{Category} & \textbf{Probe} & \textbf{Description} \\
\midrule
  \multicolumn{3}{l}{\textit{RDP}} \\
  & \texttt{RDP X.224 (mstshash=beio)} & X.224 Connection Request with routing token \texttt{mstshash=beio}; cookie value is the scanner's fingerprint \\
  & \texttt{RDP X.224 bare CR} & Minimal X.224 Connection Request with no routing cookie; detects RDP servers that accept connections without a username hint \\
  & \texttt{RDP X.224 (mstshash=Administrator)} & Same X.224 TPDU with \texttt{mstshash=Administrator}, simulating a login hint from the built-in admin account \\
\addlinespace[3pt]
  \multicolumn{3}{l}{\textit{TLS}} \\
  & \texttt{TLS 1.0 ClientHello (v1)} & TLS 1.0 ClientHello with a fixed cipher suite list; detects any TLS-capable service \\
  & \texttt{TLS 1.0 ClientHello (v2)} & Alternate TLS 1.0 ClientHello with a different cipher suite list; covers servers that reject the first hello \\
\addlinespace[3pt]
  \multicolumn{3}{l}{\textit{HTTP/Web/SIP}} \\
  & \texttt{HTTP GET /} & Plain \texttt{GET / HTTP/1.0}; baseline probe to detect any HTTP server and retrieve response headers \\
  & \texttt{HTTP OPTIONS /} & \texttt{OPTIONS / HTTP/1.1}; enumerates allowed methods and identifies HTTP servers that reject GET \\
  & \texttt{RTSP OPTIONS /} & \texttt{OPTIONS / RTSP/1.0}; detects streaming media servers and network-attached cameras/DVRs \\
  & \texttt{HTTP Nmap probe} & Nmap HTTP probe (\texttt{GET /nice\%20ports\%2C/Tri\%6Eity.txt\%2ebak"}); unusual URL elicits a clean error response for fingerprinting \\
  & \texttt{SIP OPTIONS} & SIP OPTIONS request; detects VoIP servers, PBX systems, and SIP proxies \\
\addlinespace[3pt]
  \multicolumn{3}{l}{\textit{Windows/AD}} \\
  & \texttt{Kerberos AS-REQ (krbtgt)} & Kerberos AS-REQ for the \texttt{krbtgt} service principal; identifies Active Directory domain controllers \\
  & \texttt{SMB1 Negotiate} & SMB1 Negotiate Protocol Request (\texttt{\textbackslash{}xffSMB}); detects Windows file shares \\
  & \texttt{MMS (Windows Media Services)} & Microsoft Media Server (MMS) protocol initiation; detects Windows Media Services streaming servers \\
  & \texttt{MS-SQL pre-login} & TDS pre-login packet; detects Microsoft SQL Server instances and elicits version information \\
  & \texttt{Microsoft MSMQ (MS-MQQB)} & Microsoft Message Queuing binary protocol session initiation; detects MSMQ queue managers (port 1801) \\
\addlinespace[3pt]
  \multicolumn{3}{l}{\textit{Database}} \\
  & \texttt{LDAP SearchRequest} & LDAP SearchRequest for the root DSE; detects LDAP directory servers and Active Directory on port 389 \\
  & \texttt{LDAP Anonymous Bind Request} & LDAP BindRequest (BER tag \texttt{0x60}) with version 2, null DN, and empty password; detects LDAP servers that permit anonymous access \\
  & \texttt{Java RMI} & Java Remote Method Invocation stream header (\texttt{JRMI}); detects Java RMI registries and servers on port 1099 \\
  & \texttt{Oracle TNS Connect} & Oracle Transparent Network Substrate Connect packet; detects Oracle database listeners, typically on port 1521 \\
  & \texttt{CORBA IIOP} & CORBA GIOP/IIOP request header; detects CORBA middleware object brokers \\
\addlinespace[3pt]
  \multicolumn{3}{l}{\textit{Network}} \\
  & \texttt{Sun RPC portmapper} & Sun RPC NULL call to the portmapper (port 111); enumerates registered RPC services including NFS and NIS \\
  & \texttt{DNS CHAOS version.bind} & DNS TXT query for \texttt{version.bind} in the CHAOS class; extracts software version strings from DNS servers \\
  & \texttt{DNS Status Request} & DNS STATUS query (opcode 2) over TCP; elicits a status response from DNS servers to confirm reachability \\
  & \texttt{AFP DSIGetStatus} & Apple Filing Protocol DSI GetStatus request (command 3); detects AFP file servers (port 548) and retrieves server capabilities \\
  & \texttt{OpenVPN TCP} & OpenVPN TCP-mode session initiation (P\_CONTROL\_HARD\_RESET\_CLIENT\_V2); detects OpenVPN servers running in TCP mode \\
  & \texttt{X11 Display Connection Setup} & X Window System connection request with little-endian byte order and no authentication data; detects X11 display servers (port 6000+) \\
  & \texttt{NFS/RPC getport} & RPC GETPORT call for the NFS program number; identifies NFS server endpoints via the portmapper \\
\addlinespace[3pt]
  \multicolumn{3}{l}{\textit{Specialty}} \\
  & \texttt{TNMP (Thomson/Trend Micro)} & Proprietary management protocol probe (\texttt{TNMP} magic bytes); detects Thomson residential gateways or Trend Micro management agents \\
  & \texttt{Tibco Rendezvous} & Tibco Rendezvous messaging daemon handshake (\texttt{DmdT} magic bytes); detects Tibco RV middleware used in financial messaging \\
  & \texttt{Lotus Notes RPC} & Lotus Notes Remote Procedure Call connection initiation; detects IBM/HCL Domino mail and groupware servers (port 1352) \\
  & \texttt{iSCSI Login Request} & iSCSI Login Request PDU; detects iSCSI storage targets (port 3260) exposing block storage over TCP \\
  & \texttt{DICOM A-ASSOCIATE-RQ} & DICOM Association Request; detects medical imaging equipment and PACS servers (port 104) \\
\addlinespace[3pt]
  \multicolumn{3}{l}{\textit{Generic}} \\
  & \texttt{Bare CRLF (banner grab)} & Sends \texttt{\textbackslash{}r\textbackslash{}n\textbackslash{}r\textbackslash{}n}; services such as SMTP, FTP, and IMAP emit a greeting banner on connect without waiting for input \\
  & \texttt{HELP banner grab} & Sends \texttt{HELP\textbackslash{}r\textbackslash{}n}; SMTP and POP3 servers respond with a list of supported commands, revealing service identity \\
  & \texttt{`.default.' probe} & Sends \texttt{\textbackslash{}x01.default.}; targets embedded device management protocols that use this magic string for service discovery \\
  & \texttt{Null probe} & Sends null bytes; a catch-all that elicits responses from custom or unknown TCP services that respond to any data \\
  & \texttt{`.I20100.' probe} & Sends \texttt{\textbackslash{}x01I20100\textbackslash{}n}; targets specific embedded or industrial device protocols that recognize this magic byte sequence \\
\addlinespace[3pt]
  \multicolumn{3}{l}{\textit{Unknown}} \\
  & \texttt{Unknown binary (156B, +{<}M magic)} & 156-byte binary payload containing a \texttt{{<}M} byte sequence at offset 6; protocol not identified \\
\bottomrule
\end{tabularx}
\end{table*}

\begin{acronym}
  \acro{AS}{Autonomous System}
  \acrodefplural{AS}[ASes]{Autonomous Systems}
  \acro{ASN}{\ac{AS} Number}
  \acro{AUP}{Acceptable Use Policy}
  \acro{BGP}{Border Gateway Protocol}
  \acro{CCPA}{California Consumer Privacy Act}
  \acro{CDN}{Content Distribution Network}
  \acro{CPE}{Customer Premises Equipment}
  \acro{CSIRT}{Computer Security Incident Response Team}
  \acro{DAD}{Duplicate Address Detection}
  \acro{EUI}{Extended Unique Identifier}
  \acro{GDPR}{General Data Protection Regulation}
  \acro{IoT}{Internet of Things}
  \acro{ISP}{Internet Service Provider}
  \acro{IID}{Interface Identifier}
  \acro{LAN}{Local Area Network}
  \acro{NIC}{Network Interface Card}
  \acro{NTP}{Network Time Protocol}
  \acro{MAC}{Media Access Control}
  \acro{OS}{Operating System}
  \acro{OUI}{Organizationally Unique Identifier}
  \acro{SOHO}{Small Office-Home Office}
  \acro{U/L}{Universal/Local}
  \acro{SLAAC}{Stateless Address Autoconfiguration}
  \acro{VPS}{Virtual Private Server}
\end{acronym}

\end{document}